\documentclass[5p]{elsarticle}%[review,number,sort&compress]

\usepackage{bm}        % for math                                   
\usepackage{amssymb}   % for math   
\usepackage{wasysym}
\newcommand{\isotope}[2]{$^{#2}$#1}			%%						
\newcommand{\bet}{$\beta^{-}$}				%%						
\newcommand{\bplus}{$\beta^{+}$}			%%						
\newcommand{\ec}{$\kappa$ }					%%						
\newcommand{\alp}{$\alpha$}					%%				
\newcommand{\g}{$\gamma$}					%%
\newcommand{\yr}{yr$^{-1}$}					%%
\newcommand{\dy}{d$^{-1}$}					%%
\def\dt {{dead-time}}						%%
\def\hf {{half-life}}						%%
\def\dr {{decay rate}}						%%
\def\lt {{live-time}}											%%
\usepackage[bookmarks=false]{hyperref}
\hypersetup{
    pdfstartview={FitH},    % fits the width of the page to the window
    colorlinks=true,        % false: boxed links; true: colored links
    linkcolor=blue,         % color of internal links
    citecolor=blue,         % color of links to bibliography
    urlcolor=blue           % color of external links
}

\pdfoutput=1
\begin{document}

%----------------------------------------------------------------------------------------
%	FRONTMATTER CONTENTS
%----------------------------------------------------------------------------------------
\begin{frontmatter}
	\title{Phenomenology of Rate--Related Nonlinear Effects in Nuclear Spectroscopy
	}

	\author[PU]{J.\ Nistor}
	\author[PU]{J.\ Heim} 
	\author[PU]{E.\ Fischbach\corref{EF}}
		\ead{ephraim@purdue.edu}
		\cortext[EF]{Corresponding author}

	\author[TAM]{J.H.\ Jenkins}
	\author[SU]{P.A.\ Sturrock}

	\address[PU]{Department of Physics and Astronomy, Purdue University, West Lafayette, IN 47906, USA}

	\address[TAM]{Department of Nuclear Engineering, Texas A{\&}M University, College Station, TX 77843, USA}

	\address[SU]{Center for Space Science and Astrophysics, Stanford University, Stanford, CA 94305, USA}

	%------------------------------------------------------------------------------------
	%	ABSTRACT CONTENTS
	%------------------------------------------------------------------------------------
	\begin{abstract}
A series of recent reports suggest that the decay rates of several isotopes may have been influenced by solar activity (perhaps by solar neutrinos). A mechanism in which neutrinos or antineutrinos can influence the decay process suggests that a sample of decaying nuclei emitting $\bar{\nu}_e$ could affect its own rate of decay. Past experiments have searched for this ``self--induced decay'' (SID) effect by measuring deviations from the expected decay rate for highly active samples of varying geometries. Here, we show that the SID effect closely resembles the behavior of rate-related losses due to \dt{}, and hence that standard \dt{} corrections can result in the removal of any SID-related behavior. We conclude by describing experiments which could disentangle SID effects from those arising from \dt{}.
	\end{abstract}

	\begin{keyword}
SID, Beta decay, Neutrinos, Nuclear decay lifetimes, Dead-time
	\end{keyword}

\end{frontmatter}
%----------------------------------------------------------------------------------------
%----------------------------------------------------------------------------------------
%----------------------------------------------------------------------------------------

%----------------------------------------------------------------------------------------
%	SECTION CONTENTS
%----------------------------------------------------------------------------------------
%----------------------------------------------------------------------------------------
%	SECTION CONTENTS
%----------------------------------------------------------------------------------------
\section{\label{sec:Introduction}Introduction}

Recently a number of groups have reported unexplained periodic variations in measured nuclear decay rates from experiments utilizing a variety of detector types and isotopes (see Table~\ref{tab:isotopes}). The common feature of the observed anomalies is that they appear to be associated with the Sun, as in the case of an annual signal presumed to arise from the annual variation of the Earth--Sun distance~\cite{three,four}, a (10--15) yr$^{-1}$ variation associated with the Sun's rotation~\cite{five}, a Rieger periodicity~\cite{six}, and a short-term statistically significant change in the \isotope{Mn}{54} decay rate coincident with a solar flare~\cite{flare}. A possible mechanism to account for a solar influence would be a coupling of the decaying system to solar neutrinos ($\nu_{\astrosun}$) via some as yet unknown interaction.

%------------------------------------------------

A test of the solar neutrino hypothesis was carried out in Refs.~\cite{Lindstrom1} and \cite{Lindstrom2}, which compared the decay rates of \isotope{Au}{198} in a thin gold foil, a gold wire, and a gold sphere having both the same mass (1 mg) and the same specific activity. The basis for this comparison was the observation that the electron-antineutrinos ($\bar{\nu}_e$) produced from those \isotope{Au}{198} atoms undergoing decay would bathe the undecayed atoms in the sphere in a flux of $\bar{\nu}_e$ that could be comparable to (or even greater than) ambient $\nu_{\astrosun}$ flux. If the effects of $\bar{\nu}_e$ on a decaying atom were similar to those of $\nu_{\astrosun}$, then the \dr{} of the spherical sample could be measurably different from that of the foil or wire, in which most $\bar{\nu}_e$ would presumably leave the sample without significant influence on the decay process. It is clear that this hypothesized ``self--induced decay'' (SID) effect in the sphere is a non-linear phenomenon since its effect on the activity of the sample depends on the activity itself. As shown in  Ref.~\cite{Lindstrom1}, the SID effect is characterized (to first order) by the differential equation 
\begin{equation}\label{eq:SIDdiff}
-\frac{dN(t)}{dt} \equiv -\dot{N}\cong \lambda_0 N(t) \left[ 1 + \xi \frac{N(t)}{N_0} \right]\;,
\end{equation} 
where $\lambda_0$ is the decay constant in the absence of SID (i.e. $\xi = 0$), $N$ is the number of activated nuclei, and $N_0 \equiv N(t=0)$. 

%------------------------------------------------

The objective of the present paper is to highlight the similarities of predicted SID behavior to those of typical \dt{} effects, which are a consequence of the fact that detector \dt{} effects are also non--linear in the activity of the sample (or more precisely the count rate). Hence, \dt{} effects could be confused with the SID effect and vice versa. The result is that certain \dt{} corrections may remove non--linearity in decay data which may arise from fundamentally physical (i.e. internal to the sample), rather than instrumental, origins.

%----------------------------------------------------------------------------------------
%	TABLE 1
\begin{table*}
\caption{\label{tab:isotopes}Various experiments where time-dependent nuclear decay rates have been observed. For each entry the observed nuclides and their dominant decay modes are exhibited. Observed periodicities in the decay rates are noted.}

\centering
\scriptsize
\resizebox{\linewidth}{!}{%
\begin{tabular}{cclclc}\hline
Isotope & Decay Type & Detector Type & Radiation Measured & Effect/Periodicity Observed & Reference \\ \hline 

\isotope{H}{3} & \bet & Photodiodes & \bet & 1 \yr & \cite{table35}\\
\isotope{H}{3} & \bet & Liquid Scintillator & \bet & 1 \dy, 12.1 \yr, 1 \yr & \cite{table36}\\
\isotope{H}{3} & \bet & Liquid Scintillator & \bet & $\sim 12.5$ \yr & \cite{table37}\\
\isotope{H}{3} & \bet & Solid State (Si) & \bet & $\sim 2$ \yr & \cite{table38}\\
\isotope{Na}{22}/\isotope{Ti}{44} & \bplus, \ec & Solid State (Ge) & \g & 1 \yr & \cite{table39}\\
\isotope{Cl}{36} & \bet & Proportional & \bet & 1 \yr, 11.7 \yr, 2.1 \yr & \cite{table3,table16,table20}\\
\isotope{Cl}{36} & \bet & Geiger--M\"{u}ller & \bet & 1 \yr & \cite{table40}\\
\isotope{Mn}{54} & \ec & Scintillation & \g & Solar flare & \cite{table15}\\
\isotope{Mn}{54} & \ec & Scintillation & \g & 1 \yr & \cite{table41}\\
\isotope{Mn}{54} & \bet & Scintillation & \g & 1 \yr & \cite{table9}\\
\isotope{Co}{60} & \bet & Geiger--M\"{u}ller & \bet , \g & 1 \yr & \cite{table5,table6}\\
\isotope{Co}{60} & \bet & Scintillation & \g & 1 \dy , 12.1 \yr & \cite{table42}\\
\isotope{Kr}{85} & \bet & Ion Chamber & \g & 1 \yr & \cite{table18}\\
\isotope{Sr}{90}/\isotope{Y}{90} & \bet & Geiger--M\"{u}ller & \bet & 1 \yr , 11.7 \yr & \cite{table5,table6,table43}\\
\isotope{Ag}{108\mathrm{m}} & \ec & Ion Chamber & \g & 1 \yr & \cite{table18}\\
\isotope{Ba}{133} & \bet & Ion Chamber & \g & 1 \yr & \cite{thiswork}\\
\isotope{Cs}{137} & \bet & Scintillation & \g & 1 \dy , 12.1 \yr & \cite{table42}\\
\isotope{Eu}{152} & \bet , \ec & Solid State (Ge) & \g  & 1 \yr & \cite{table33}\\
\isotope{Eu}{152} & \bet , \ec & Ion Chamber & \g & 1 \yr & \cite{table18}\\
\isotope{Eu}{154} & \bet , \ec & Ion Chamber & \g & 1 \yr & \cite{table18}\\
\isotope{Rn}{222} & \alp , \bet & Scintillation & \g & 1 \yr , 11.7 \yr , 2.1 \yr & \cite{table44,table45}\\
\isotope{Ra}{226} & \alp , \bet & Ion Chamber & \g & 1 \yr , 11.7 \yr , 2.1 \yr & \cite{table3,five,table20}\\
\isotope{Pu}{239} & \bet & Solid State & \alp & 1 \dy , 13.5 \yr , 1 \yr & \cite{table36}\\
\hline
\end{tabular}
}
\end{table*}
%----------------------------------------------------------------------------------------

%----------------------------------------------------------------------------------------
%	SECTION END
%----------------------------------------------------------------------------------------
%----------------------------------------------------------------------------------------
%	SECTION CONTENTS
%----------------------------------------------------------------------------------------
\section{Experimental Motivation}\label{sec:two}

%----------------------------------------------------------------------------------------
%	TABLE TWO
%----------------------------------------------------------------------------------------
\begingroup
\begin{table}[b]
\caption{\label{LS_SID_table}Exponential and SID fits to net counts for various \isotope{Au}{198} data sets. The NIST published value for \isotope{Au}{198} is $T_{1/2} = 64.684 \pm 0.005 $ hr.}
\centering
\small
\resizebox{\columnwidth}{!}{%
	\begin{tabular}{ccccc}\hline
		Run & $\displaystyle{T_{1/2\; \mathrm{exp}}}$ (hr) & $\displaystyle{\chi^2 _{\;\mathrm{exp}}}$ & $\displaystyle{T_{1/2\; \mathrm{SID}}}$ (hr) & $\displaystyle{\chi^2 _{\;\mathrm{SID}}}$ \\ \hline
		
		Au105 & 64.365 $\pm$ 0.007 & 1.44 & 64.642 $\pm$ 0.008 & 0.98 \\
		Au301 & 64.062 $\pm$ 0.017 & 1.52 & 64.629 $\pm$ 0.019 & 1.04 \\
		Au401 & 64.394 $\pm$ 0.005 & 1.56 & 64.610 $\pm$ 0.006 & 1.12 \\
		Au501 & 64.484 $\pm$ 0.007 & 1.10 & 64.574 $\pm$ 0.008 & 1.05 \\
		Au502 & 64.078 $\pm$ 0.004 & 6.32 & 64.672 $\pm$ 0.005 & 1.05 \\
		\hline
	\end{tabular}
	}
\end{table}
\endgroup

A seemingly unrelated experiment was conducted by our group, which provided the motivation for revisiting the SID prescription. In this experiment, several short-lived isotopes were irradiated on a regular basis and subsequently observed in order to measure the half-lives repeatedly throughout the year. The primary objective was to determine whether shorter lived isotopes also exhibit an annual time-dependence in their decay rates, similar to the isotopes presented in Table~\ref{tab:isotopes}, which would otherwise be obscured by their short life-times. 

%------------------------------------------------

Of interest to this discussion are the data collected on  \isotope{Au}{198}, which were in the form of thin foils, typically of mass 12-13 mg. The foils were placed in labeled polypropylene vials and heat-sealed. On a weekly basis, as possible, a sample was irradiated for 3 to 5 minutes in Purdue's 1 kW research reactor (PUR1), at a flux of $\sim 1 \times 10^9$ n/cm$^2$s. Immediately after irradiation, the sample was moved to a 3-inch Bicron NaI(Tl) well detector, where consecutive 900--second \lt{} collection periods were recorded for 2 to 4 half-lives. Initial count rates were typically approximately 10 kcps, keeping \dt{} below 10\%. Data were collected using an Ortec Digibase and the Maestro-32 MCA software package. A Region of Interest (ROI) was set around the 412 keV \g--peak, and a peak-stabilization feature within Maestro was enabled. 

%------------------------------------------------

\begin{figure*}
\centering
\includegraphics[width=1\linewidth]{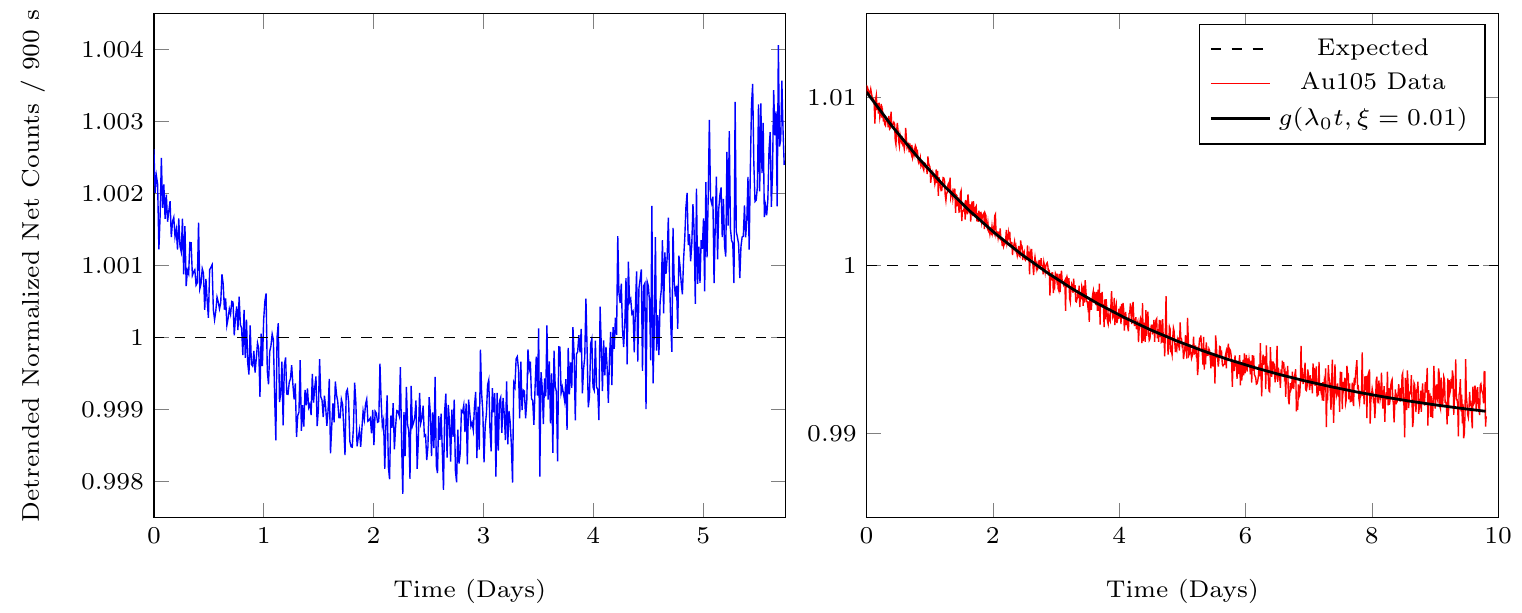}
\caption{Detrended and normalized \isotope{Au}{198} counts per 900 second fixed \lt{} interval. ({\em Left}) The data were detrended assuming a pure exponential decay model. ({\em Right}) A representative example of the detrended \isotope{Au}{198} net counts (red, solid curve) using best-fit SID parameters. The expected behavior in the absence of SID or experimental bias (black, dashed line) is a random distribution of points about unity. The prediction from a SID perturbation is represented by the solid (black) curve. The function $g(\lambda_0 t, \xi)$ is defined in Eq.\ (\ref{eq:g}).}\label{fig1}
\end{figure*}

The integral data from the dominant 412 keV peak were analyzed using a weighted least-squares fit to an exponential decay law. Half-life determinations were near the expected value ($T_{1/2} = 64.684(5)$ hr), but exhibited suspect variations within subsets of the data, and for data sets with significantly higher count rates. Further analysis repeatedly showed an anomaly in the residuals of the detrended count rates, as depicted in Fig.~\ref{fig1}, which are expected to be distributed normally about unity.
%
%----------------------------------------------------------------------------------------
%	FIGURE ONE
%----------------------------------------------------------------------------------------
Similar anomalies were seen in data from other nuclides measured with different detection systems. While investigating the role of \dt{} in this behavior, it was discovered that a SID-corrected fit matched the data extremely well, and yielded a more consistent result for \hf{} determinations for many of the data sets.

%------------------------------------------------

\begin{figure*}[t]
\centering
\includegraphics[width=.9\linewidth]{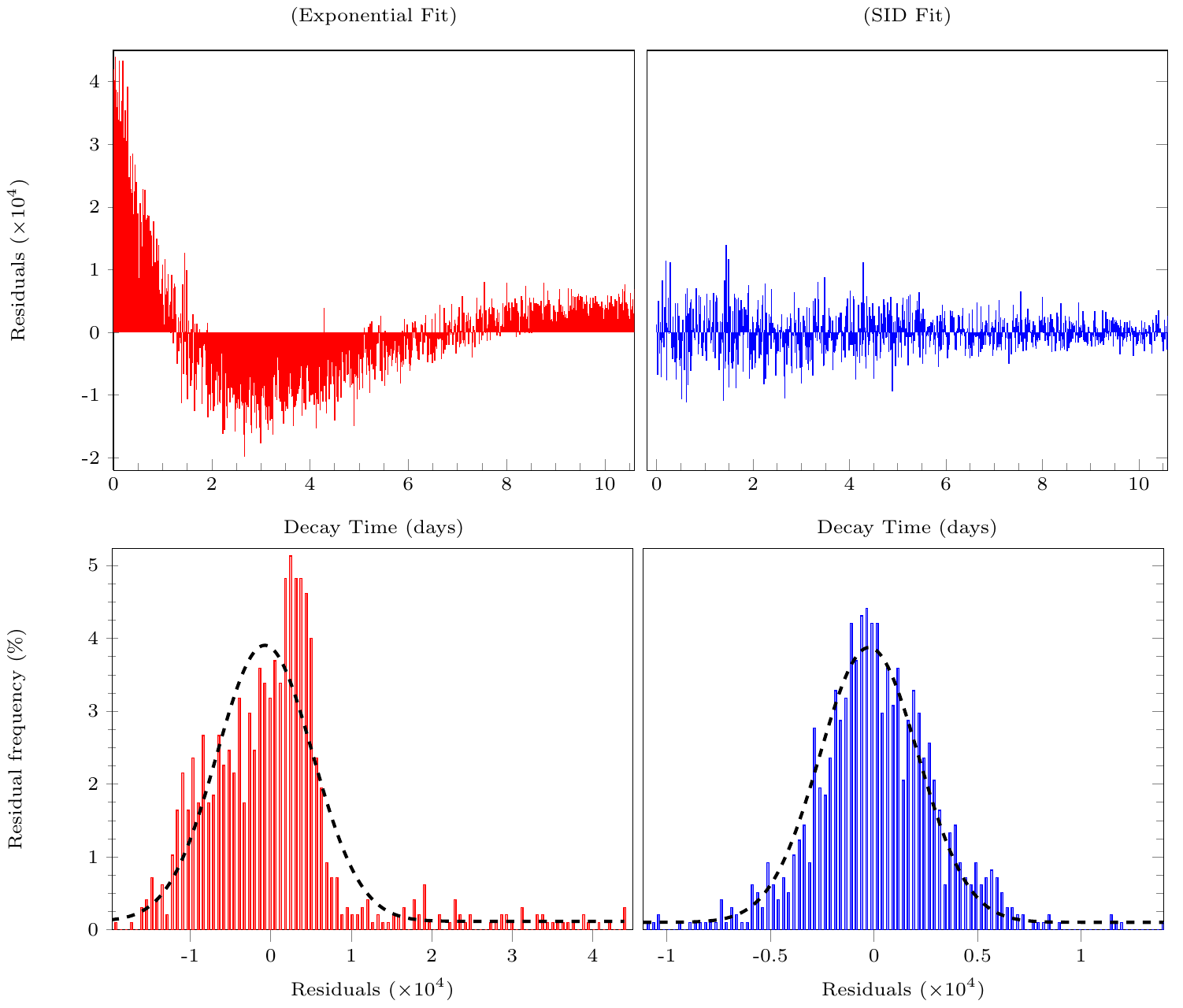}
\caption{Residuals for run Au105 are presented for both exponential ({\em left}) and SID ({\em right}) models. The residuals for the exponential model demonstrate a significant temporal bias which is absent for the SID fit. The corresponding histograms are displayed in the bottom graphs.}\label{au502_exp_hist}
\end{figure*}

To evaluate the possible presence of a SID effect in our samples, the decay data were analyzed by a weighted best-fit to the first-order SID function (see discussion in Sec.~\ref{SID_Deadtime_Comparison}). The relevant fit parameters were then used to detrend the decay data, and the results (illustrated in Fig.~\ref{au502_exp_hist}) closely matched the SID prediction presented in Ref.~\cite{Lindstrom1}.  A SID presence would initially cause depopulation of the sample more quickly than predicted by purely exponential behavior. This in turn would cause the subsequent decay rate to be smaller than expected under exponential behavior, and this crossover occurs at $T_{1/2}$.

%-----------------------------------------------

Calculated values for $\chi^2$ per degree of freedom, $\displaystyle{\chi^2 _{\mathrm{DOF}}}$, presented in Table \ref{LS_SID_table}, show improvement for the SID prescription compared to an exponential fit. Of particular note is run `Au502', whose initial activity was roughly double that of the next most active \isotope{Au}{198} sample observed. A higher specific activity would be expected to lead to a stronger SID effect. While the SID prescription continues to fit very well with this larger non-exponential behavior, the purely exponential fit is poor, yielding $\chi^2 _{\mathrm{DOF}} \approx 6$. It is also true that a larger initial activity corresponds to a higher count rate, and hence, increased \dt{} in our experiment. While the Gedcke-Hale \lt{} clock utilized in this experiment corrects for rate-related losses such as \dt{} and pileup, there exists an uncertainty in each time-correction. A systematic over/under correction would directly affect an accurate determination of the SID parameter, $\xi$. Further experimentation may offer a way to distinguish these two effects, as we discuss below.

%-----------------------------------------------

Analysis of the residuals of the net counts for both the purely exponential fit and the SID fit highlights the differences between the two methods of analysis presented here. The histogram of residuals for the SID fit, presented in Fig. \ref{au502_exp_hist}, shows an approximately Gaussian distribution, whereas the corresponding distribution for the exponential fit is skewed. Plots of the residuals as a function of time reveal further differences; under the SID prescription, the residuals exhibit a relatively normal distribution about zero, whereas the residuals of the exponential fit show a marked departure. The shape of the exponential-fit residuals is consistent with the sample initially decaying faster than predicted by an exponential model. Analysis of data for other nuclides also shows improvements in fit under SID analysis. Table~\ref{avg_table} displays the averages of the determined \hf{} for each nuclide, weighted by the associated error for each datum in the determination. The average $\displaystyle{\chi^2 _{\mathrm{DOF}}}$ for the exponential and SID fits is also presented, including the range. In all cases, the SID model is a better description of the data than a pure exponential model.

%----------------------------------------------------------------------------------------
%	TABLE THREE
%----------------------------------------------------------------------------------------
\begingroup
\begin{table*}[t]
\caption{\label{avg_table}Weighted averages of the half-lives of various isotopes using exponential and SID fits to net counts. The associated mean $\chi^2$ per degree of freedom and range are presented.}

\centering
\scriptsize
\resizebox{\linewidth}{!}{%
		\begin{tabular}{lllllll}\hline
Isotope & $T_{1/2\; \mathrm{exp}}$ (days) & Avg $\chi^2 _{\mathrm{exp}}$ & (min, max) & $T_{1/2\; \mathrm{SID}}$ (days) & Avg $\chi^2 _{\mathrm{SID}}$ & (min, max) \\ \hline

\isotope{Au}{198} & 2.68426 $\pm$ 3.9$\times 10^{-5}$ & 2.39 &(1.10, 6.32) & 2.6929 $\pm$ 1.4$\times 10^{-4}$ & 1.05 &(0.98, 1.12)\\
\isotope{As}{76} & 1.09159 $\pm$ 8.7$\times 10^{-5}$ & 1.17 &(1.03, 1.29) & 1.0906 $\pm$ 3.6$\times 10^{-4}$ & 1.14 &(1.00, 1.24)\\
\isotope{Sb}{122} & 2.67735 $\pm$ 9.5$\times 10^{-5}$ & 5.30 &(1.72, 12.3) & 2.6531 $\pm$ 3.5$\times 10^{-4}$ & 2.06 &(1.14, 3.33)\\
\isotope{Mn}{56} & 0.10690 $\pm$ 5.9$\times 10^{-5}$ & 8.38 &(1.14, 24.5) & 0.10755 $\pm$ 1.26$\times 10^{-5}$ & 1.19 &(0.78, 2.01)\\
\isotope{In}{116m} & 0.037906 $\pm$ 9.3$\times 10^{-6}$ & 1.83 &(0.97, 2.17) & 0.03763 $\pm$ 1.8$\times 10^{-5}$ & 1.25 &(0.95, 2.14)\\
\hline
		\end{tabular}
}
\end{table*}
\endgroup

%----------------------------------------------------------------------------------------
%	SECTION END
%----------------------------------------------------------------------------------------

%----------------------------------------------------------------------------------------
%	SECTION CONTENTS
%----------------------------------------------------------------------------------------
\section{SID Phenomenology}\label{SID_Deadtime_Comparison}

On a phenomenological basis, a natural way to investigate the periodicities presented in Table\ \ref{tab:isotopes}, as well as a possible influence from neutrinos, is to model the behavior as a modification to the standard exponential decay law, $-dN/dt = \lambda N$, where the decay parameter $\lambda$ experiences a time--dependent perturbation,\footnote{We note that a time--dependence in $\lambda$ does not necessarily imply a departure from randomness, but rather suggests a deviation in the probability distribution which governs the decay.} i.e.,
\begin{equation}
\lambda = \lambda_0 + \lambda_1 (t)\;.
\end{equation} 
Since the perturbation is presumed to arise from an interaction with neutrinos or antineutrinos (which will henceforth be referred to as neutrinos), $\lambda_1$ will be proportional to (or more generally, a function of) the ambient neutrino flux, which in principle may have contributions from a variety of sources (e.g. solar neutrinos, C$\nu$B relic neutrinos, geologic and atmospheric neutrinos, and artificial/reactor-generated neutrinos). For such an exotic interaction to exist in nature, past experimental observations constrain $\lambda_1$ to be much less than $\lambda_0$, at least for conditions typically encountered in terrestrial experiments.\footnote{One could imagine locations where the ambient neutrino flux is significantly larger than that encountered on Earth---such as near or within stellar bodies, for instance.} This constraint suggests that it is appropriate to simplify subsequent expressions to lowest order in $\lambda_1/\lambda_0$. For the purpose of comparing SID with \dt{} behavior, higher order SID terms will be discarded. The one exception will be throughout the discussion of the ``extremal'' behavior associated with SID.

%------------------------------------------------

As mentioned in Sec.\ \ref{sec:Introduction}, it is hypothesized that a sample undergoing $\beta$--decay or $K$--capture may in fact be able to affect its own rate of decay. Specifically, those atoms which have yet to decay will be bathed in a flux of neutrinos produced from the decaying atoms within the sample. Therefore, a sample with a sufficient internal neutrino flux (i.e., greater than the $\nu_{\astrosun}$ flux) should exhibit an experimentally detectable deviation in its decay rate. It is perhaps surprising that an internal neutrino flux significantly higher than the $\nu_{\astrosun}$ flux is achievable in relatively small samples, as was accomplished in \cite{Lindstrom1,Lindstrom2} with a 1 mg gold sphere, foil, and wire.

%------------------------------------------------

Given such a case, where the various contributions to the total neutrino flux (within a given sample) are dominated by the the internally-generated neutrino flux, it is reasonable to neglect all but this internal source and express the perturbation in the decay parameter as
\begin{equation}\label{eq:perturb}
%\lambda_1 = -\xi\left(\frac{\dot{N}}{N_0}\right)  .
\lambda_1 = -p \dot{N} .
\end{equation}
Explicitly, it is presumed that the perturbation is proportional to the density of emitted neutrinos, which in turn is proportional to the decay rate ($p$ being the dimensionless proportionality constant). To be more general than Eq.\ (\ref{eq:perturb}), the decay parameter may contain additional perturbations arising from external neutrinos (which is presumed to be the case for the periodicities observed in Table\ \ref{tab:isotopes}). However, for the specific case of a short-lived isotope, one would expect the time-dependence in $\lambda$ to arise primarily from $\dot{N}$, while the $\nu_{\astrosun}$ contribution, for instance, would remain approximately constant during the sample's short lifetime. Using Eq.\ (\ref{eq:perturb}), the perturbed system can be characterized by the differential equation:
\begin{eqnarray}
-\dot{N}&=& \left(\lambda_0  - p \dot{N}\right)  N  \label{eq:SIDdiff_exact}\\
&=& \frac{\lambda_0 N}{1-p N}\;.\label{eq:SIDdiff_exact1}
\end{eqnarray}

%----------------------------------------------------------------------------------------
%	FIGURE TWO
%----------------------------------------------------------------------------------------

%------------------------------------------------

Before presenting a solution to Eq.\ (\ref{eq:SIDdiff_exact1}), which is relevant for comparison with \dt{} effects, we highlight a few points which have not been previously presented elsewhere. Since $\lambda dt$ can be regarded as the probability for an unstable atom to decay within a small time interval $dt$, it is reasonable to interpret $\lambda_1 dt$ as the probability within this time for an activated atom to decay, due to the proposed SID effect. In other words, the probability for an atom to decay is the sum of two distinct ways in which the decay can occur: a {\em spontaneous} decay ($\lambda_0 dt$), or an {\em induced} decay ($\lambda_1 dt$). In this context, the constant of proportionality in Eq.\ (\ref{eq:perturb}) can be regarded as the probability (per emitted neutrino) for a SID to occur, since $-\dot{N} dt$ is the approximate number of neutrinos emitted in time $dt$.\footnote{It may be that $p$ is negative, in which case, $-p$ is the probability of inhibiting a decay event from occurring. The treatment that follows will assume a positive value for $p$, but the treatment for negative $p$ is readily inferred.} Therefore, each neutrino emitted within a population of $N$ activated atoms is expected to induce $pN$ decays. It may appear at first glance that $pN$ could exceed unity if a single neutrino were able to influence multiple unstable atoms while traveling within the sample. In fact, one could in principle attempt to irradiate a sample continuously until $N$ becomes large enough so that $pN$ is greater than unity; however, the technical question is how such a sample would be created.

%------------------------------------------------

\begin{figure}[t]
\includegraphics[width=.9\columnwidth]{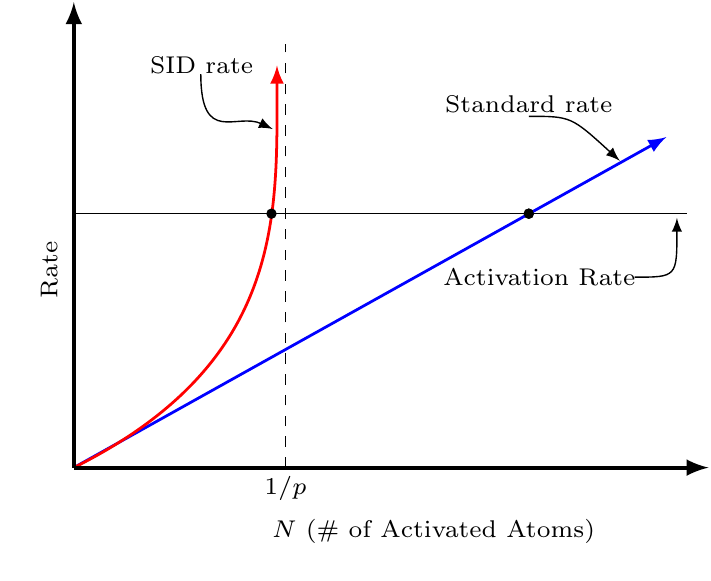}
\caption{An illustration depicting {\em secular equilibrium} for a sample exhibiting the SID behavior (red curve) and a sample undergoing standard exponential decay (blue curve). For a given activation rate, equilibrium is reached with {\em fewer} activated atoms for a sample undergoing SID decay, and the absolute maximum number of activated atoms is $N_{\mathrm{max}} = 1 /p$.}\label{fig:cartoon}
\end{figure}

To address this question, first consider an ``ideal'' sample of identical stable atoms which can be irradiated at will (perhaps via neutron activation) by placing the sample in a reactor, for example. Since the activated atoms in the sample will subsequently decay, there is a limit to how many stable atoms become activated---a limit which is reached when the decay rate of the sample (which increases as more atoms are activated) equals the activation rate of the reactor. If the sample being considered decays according to Eq.\ (\ref{eq:SIDdiff_exact1}), then as $pN$ approaches unity the decay rate diverges. No matter how powerful the reactor or how large the sample is, the activation rate will be insufficient to activate more than $1/p$ atoms within the sample, i.e,
\begin{equation}
N_{\mathrm{max}} = 1/p\;.
\end{equation}
This new example of secular equilibrium is depicted in Fig.\ \ref{fig:cartoon}, and is in stark contrast to a sample governed by the standard decay law, for which the absolute maximum number of activated atoms is only limited to the total number of atoms within the population. From this discussion, there is an unavoidable upper bound on $pN$ such that:
\begin{equation}\label{eq:pbound}
\xi \equiv pN_0  < 1\;,
\end{equation}
where $N_0 < N_{\mathrm{max}}$ is the number of activated atoms immediately after the activation process is completed. Note that the SID parameter $\xi$ is the quantity which may be directly measurable by experiment, since it represents the fractional change in the decay rate from the expected behavior at $t = 0$.

%----------------------------------------------------------------------------------------
%	FIGURE THREE
%----------------------------------------------------------------------------------------

%------------------------------------------------

Finally, given the aforementioned interpretation of $p$, we could have derived Eq.\ (\ref{eq:SIDdiff_exact1}) simply by adding up all of the decays expected within a short time interval $dt$ through the distinct ways a decay can occur. Since there are $\lambda_0 N dt$ decays expected to occur randomly, there will also be this many neutrinos available to stimulate additional events. With $pN$ being the number of SID events expected per neutrino, $\lambda_0 N dt (pN)$ would be the expected number of induced decays. However, these SID decays, in turn, produce additional neutrinos for which $\lambda_0 N dt (pN)^2$ decays are expected. The pattern continues since at each stage the additional decays in turn produce additional neutrinos. Therefore, we can express the total number of expected decays, per unit time, as
\begin{equation}\label{eq:sidexpand}
-\dot{N} = \lambda_0 N  + p \lambda_0 N^2 + ... = \lambda_0 N \sum_{j=0}^{\infty} \left(pN\right)^j,
\end{equation}
where the right side of Eq.\ (\ref{eq:sidexpand}) will always converge for $|\xi| < 1$, and is readily identified with Eq.\ (\ref{eq:SIDdiff_exact1}). From this construction, it is evident that ``higher--order SID processes'' will have little influence on the total decay rate of the sample, given that $|\xi|$ is presumably much smaller than unity. Furthermore, Eq.\ (\ref{eq:SIDdiff}) can be obtained by retaining only the first two terms in Eq.\ (\ref{eq:sidexpand}).

%------------------------------------------------

The exact solution to Eq.~(\ref{eq:SIDdiff_exact1}) is most likely never required. Nevertheless, an implicit solution is obtained by integration, yielding
\begin{equation}\label{eq:sidrel}
N(t) e^{-\xi N(t)/N_0} = N_0 e^{-\lambda_0 t} e^{-\xi}.
\end{equation}
We can expand Eq.\ (\ref{eq:sidrel}) in (pseudo) powers of $\xi$ by making use of the {\em Lagrange inversion theorem}, 
\begin{equation}\label{eq:powers}
N(t) = \frac{N_0}{\xi} \sum_{j=1}^{\infty} \frac{j^{j-1}}{j!} \left( \xi e^{-\xi} e^{-\lambda_0 t} \right)^j ,
\end{equation}
where $\xi$ is assured to fall within the radius of convergence on purely physical grounds. However, it is worth emphasizing that Eq.\ (\ref{eq:powers}) is an accurate solution for situations up to $\xi =1$, and therefore may be useful in studying the extremal behavior following from the preceding discussion. The decay rate, $n_{\xi}(t) \equiv -dN/dt$, can be determined from Eq.\ (\ref{eq:powers}) as a series expansion. For typical situations it is sufficient to truncate the expansion to first order in $\xi$, in which case the decay rate becomes 
\begin{equation}
\label{eq:nsid}
n_{\xi}(t) = \frac{n_{\xi}(0)}{1+2\xi} e^{-\lambda_0 t}\left(1+2\xi e^{-\lambda_0 t}\right),
\end{equation}
where $n_\xi (0)$ is the initial decay rate, and the truncation error is on the order of $\mathcal{O}(\xi^2)$. In general, terminating the expansion to the $j$th term results in a remainder $R_j (\xi,t)$ which decays much more quickly in time than the activity of the sample. In fact, this remainder is comparable to $R_j (\xi,t) \sim \left(\xi e^{1-\xi} e^{-\lambda_0 t} \right)^{j +1}$. Furthermore, we see directly from Eq.\ (\ref{eq:SIDdiff_exact1}) that the initial SID rate $n_{\xi}(0)$ is greater (when $\xi > 0$) than the initial non-perturbed rate, and the two are related according to
\begin{equation}
\left(1-\xi\right) n_{\xi} (0) = n_0 (0)\;,
\end{equation}
where $n_0 (0) \equiv \lambda_0 N_0$ is typically understood to be the initial activity of a sample.

%------------------------------------------------

In past discussions in the literature, one useful quantity that has been studied is the ratio of decay rates with and without a SID effect:
\begin{equation}\label{eq:g}
g(x,\xi) \equiv \frac{n_{\xi}}{n_0} = 1-\xi +2\xi e^{-x}\;.
\end{equation}
Previous searches for the SID effect (such as those conducted in Refs.~\cite{Lindstrom1} and \cite{Lindstrom2}) sought to compare samples of similar activities but differing geometries. In this case, the values of $\xi$ will differ between the samples due to the geometric dependence on $p$. Consequently, the ratio of the decay rates of one sample to the other is expected to have a temporal trend proportional to $g(\lambda_0 t,\Delta\xi)$, where $\Delta \xi$ is the difference between the values of $\xi$ for each sample. A plot of $g(x,\Delta\xi)$ is shown in Fig. 1 of Ref.~\cite{Lindstrom1}.

%------------------------------------------------

It will be useful in what follows to rewrite $n_\xi$ in terms of a purely exponential rate in order to compare this behavior to the perturbations associated with \dt{}, which will be discussed in Sec.\ \ref{sec:deadtime}. In fact, the {\em event rate} as seen by the front end of a detector can be written as 
\begin{equation}\label{sidrate}
m_{\xi} (t) = m_0 (t) \left[ 1 + \beta m_0 (t) \right]\;,
\end{equation}
where $\beta \equiv 2\xi/m_0 (0)$, and $m_0 (t)$ represents an exponential decay rate with decay constant $\lambda_0$. 
It is apparent from Eq.~(\ref{sidrate}) that the fractional change in the count rate $(m_{\xi}-m_0)/m_0$ shares the same time--dependence as $m_0(t)$. It is {\em also} true that rate-related perturbations associated with \dt{} losses are proportional to the count rate. Therefore, we turn now to investigate the effects of \dt{} on detector event rates and compare these effects the SID rates discussed above.
%----------------------------------------------------------------------------------------
%	SECTION CONTENTS
%----------------------------------------------------------------------------------------

\section{Dead-time behavior}\label{sec:deadtime}

We proceed in this section to a discussion of \dt{} behavior with the aim of exhibiting the similarity between \dt{} effects and those arising from the SID behavior. Rate-related losses due to \dt{} may arise from any part of the counting system, and the magnitude of these losses is directly proportional to the counting rate, itself. Many procedures exist to correct for rate-related losses. Typically these procedures are validated under conditions far more severe than those found in routine metrology, and thus are considered well-motivated. The subject of this section is to study a few simple \dt{} models in an effort to elucidate their effects on time-dependent event rates. To this end, we consider the somewhat idealized behaviors of {\em extending} and {\em non-extending} \dt{}.

%----------------------------------------------------------------------------------------
%	SUBSECTION CONTENTS
%----------------------------------------------------------------------------------------
\subsection{Extending \dt{}}\label{sec:extending}

Pileup in the amplifier (also called random summing) is a classic case of extending \dt{}. When counting rates are relatively high, the random spacing of radiation pulses may result in interfering effects between pulses. {\em Peak pileup} occurs if two pulses are sufficiently coincident in time that they are treated as a single pulse in the counting system. The effect of peak pileup of two events is to essentially shift both from their proper position in the energy spectrum. {\em Tail pileup}, which can occur significantly even at relatively low count rates, involves the superposition of two slightly overlapping pulses. The main effect of tail pileup on the measurement is to worsen and distort the spectrum resolution.

%------------------------------------------------

The counting losses which result from pileup can be modeled as those events which occur within a time spacing less than a particular time following a previous event. Let $\alpha$ denote the minimum time by which two events must be spaced in order for each event to be resolved properly, $I$ represent the instantaneous event rate of the decay within the detector, and $I'$ represent the measured event rate out of the amplifier. The fraction of events which are spaced by a time interval between $T$ and $T + dT$ is given by $I dT \exp \left(-I T\right) $, which can be interpreted as the probability of all nuclei surviving for a time $T$ followed by a decay in a time $dT$. Therefore, the fraction of events which are {{\em not}} piled up, $f$, is given by
\begin{equation}\label{eq:pileupfrac}
f = \int_{\alpha}^{\infty} e^{\displaystyle{-I T}} I dT = e^{\displaystyle{-\alpha I}} \;.
\end{equation}
Consequently, if only those events free from pileup contribute to the measured rate, then the measured rate $I'$ can be expressed as 
\begin{equation}
\label{eq:pileup}
I' = I e^{\displaystyle{-\alpha I}},
\end{equation} 
where the extending \dt{} (pileup) parameter $\alpha$ can be determined in successive counting experiments through a least-squares type fit. Pileup {\em correction factors} $f_P$, by which each datum is adjusted, take the form
\begin{equation}\label{eq:Fp}
f_P \equiv \frac{I}{I'} = e^{\displaystyle{\alpha I}}.
\end{equation}
Another representation of these correction factors is obtained by using the macroscopic \dt{} ($DT$) to approximate the counting rate, yielding
\begin{equation}
\frac{I}{I'} = e^{\displaystyle{P\left( DT/LT\right)}}\;,
\end{equation}
where $LT$ denotes the detector \lt{} for the counting period. The unknown pileup constant $P$ along with $I$ are determined from a least-squares fit of $\ln I' = \ln I - P\left( DT/LT\right)$. It is worth noting that in either case, the constants $P$ and $\alpha$ are usually not measured directly. Rather, they are best-fit determinations of a least-squares or $\chi^2$ type minimization procedure.

%------------------------------------------------

It is a curious observation that a SID--related deviation from an exponentially decaying event rate resembles the effect of pileup in the amplifier. In fact, expanding Eq.~(\ref{eq:pileup}) in powers of $\alpha$, we obtain
\begin{equation}\label{eq:pilupexpanded}
I'(t) = I(t) \left[ 1 - \alpha I \left(t\right) \right] + \mathcal{O} (\alpha^2),
\end{equation}
which bears the same form as Eq.~(\ref{sidrate}) for an exponentially decaying source. Although $\alpha$ is strictly positive to represent losses from pileup, $\xi$ can be positive or negative depending on the particular mechanism -- appearing as ``rate-related gains'' ($\xi > 0$) or rate-related losses ($\xi <0$).

%------------------------------------------------

The similarity of Eqs.\ (\ref{sidrate}) and (\ref{eq:pilupexpanded}) raises the question of what the effect from pileup would be on a SID--perturbed count rate, $m_\xi (t)$. As outlined above, the general form of Eq.~(\ref{eq:pileup}) is obtained by considering the probability for two incoming events to occur within a particular time interval. Furthermore, these events are assumed to be governed by a Poisson process, or equivalently, presumed to occur randomly in time and independent from one another. While the condition for randomness may remain valid when considering a SID process, the condition for {\em independence} falls under scrutiny. Namely, each induced decay is the result of prior events and therefore cannot be considered to occur independently. In fact, it is likely that each SID event would essentially occur almost simultaneously with the event which induced it, and therefore a disproportionately large number of events would be subjected to peak pileup.

%------------------------------------------------

The derivation of an exact expression analogous to Eq.\ (\ref{eq:pileup}) for a SID process would require a rigorous development of the distribution of time intervals for this process. An approximation suitable when $\xi$ is sufficiently small can be obtained by estimating an atom's survival probability for a time $T$ to be $N(T) / N_0$, where $N(T)$ is given by Eq.\ (\ref{eq:powers}). Additionally, the probability for an atom to decay is necessarily time--dependent, as underscored in the modified form for $\lambda(t)$. To lowest order in $\xi$, the effect from pileup on a SID event rate $m_{\xi}$, given in Eq.\ (\ref{sidrate}), can be represented as  
\begin{equation}\label{eq:stuff}
I'(t) = {m_0}(t) e^{\displaystyle{-\alpha'  {m_0}(t)}} .
\end{equation}
where $\alpha' \equiv \alpha - \beta = \alpha - 2\xi/m_0(0)$ can be considered the ``effective'' pileup parameter, and $\mathcal{O}\displaystyle{(\alpha\cdot\xi)}$ terms are considered to be of higher order. Since $m_0$ represents a pure exponential decay rate, the result obtained in Eq.~(\ref{eq:stuff}) demonstrates that a SID--modified input signal can be mistaken for an exponential decay rate (with the same decay constant $\lambda_0$) after subject to pileup. Therefore, procedures which correct each datum for pulse pileup could remove a SID perturbation almost entirely with an incorrect determination of the pileup parameter, $\alpha$, or $P$.

%------------------------------------------------

A standard hardware solution to pileup is accomplished through the use of a pulser to estimate the correction factor as the ratio of pulser frequency with and without a source. Suppose a periodic pulse generator with frequency $I_{P}$ is used to add an artificial peak to the spectrum being studied. Since the true input rate of the pulser is known, a measurement of the counts in the artificial peak allows for a determination of the fraction of events free from pileup. Once again, this fraction is given by Eq.\ (\ref{eq:pileupfrac}) for a Poisson governed decay rate, and therefore the output rate from the pulser, $I'_P$, can be expressed as
\begin{equation}\label{eq:pulserpileup}
I'_P = I_{P} e^{\displaystyle{-\alpha I}},
\end{equation}
where $I$ once again represents the true event rate due to the decay. On the other hand for a SID event rate, Eq.\ (\ref{eq:pulserpileup}) is unaffected to first order in $\xi$. This can be explained by the fact that the fraction of events which are free from pileup, $f$, is unaffected to lowest order in $\xi$. The perturbation to the output signal arises solely from the increase in the event rate. Since the pulser input rate is independent of the decay rate, the pulser method of correcting for pileup losses is immune from the removal of first-order SID effects.

%----------------------------------------------------------------------------------------
%	SUBSECTION CONTENTS
%----------------------------------------------------------------------------------------
\subsection{Non-extending \dt{}}
Non-extending \dt{} is perhaps the simplest model for a detector's response to an input signal $I$. For this model, the counting systems is ``busy'' (i.e. unable to receive any additional pulses) for a fixed time $\tau$ after each registered event. As a result, the measured counting rate, $I'$, will be an underestimate of the true event rate according to:
\begin{equation}\label{eq:nonextending}
I' = \frac{I}{1 + \tau I}.
\end{equation}

%------------------------------------------------

The total amount of time for which the detector is ``dead'' during a counting interval, called the macroscopic \dt{} $DT$, is given simply as the total accumulation of these small intervals $\tau$ for the total number of registered events. That is to say, if $M$ is the total number of registered counts in a time $CT$, then the detector is unable to register events for a time given by $DT = M  \tau$. The effective \lt{} ($LT$) counting interval, therefore, is the time $LT = CT - DT$.

%------------------------------------------------

When the true event rate into a detector is small (that is when the time between events $\sim 1/I$ is significant in comparison to the detector response time $\tau$), the non-extending \dt{} behavior given by Eq.\ (\ref{eq:nonextending}) will agree with the {\em extending} \dt{} model in Eq.\ (\ref{eq:pileup}). We can see this from expanding Eq.\ (\ref{eq:nonextending}) in powers of $\tau I$:
\begin{equation}\label{eq:deatime}
I' = \frac{I}{1+\tau I} = I \left( 1 - \tau I + \mathcal{O}\left(\tau^2 I^2\right)\right)\;.
\end{equation}
Although the two models agree for low event rates, the behaviors diverge significantly for high rates, and for the non-paralyzable model described by Eq.\ (\ref{eq:nonextending}), the observed count rate will asymptote to a value of $1/\tau$ for large, true event rates.

%------------------------------------------------

It is common practice in nuclear spectrometry to perform successive counting measurements by fixing the \lt{} clock, since the \dt{} $DT$ is expected to change during each measurement when the decaying source has a lifetime comparable to the total time through which the experiment is conducted. If, on the other hand, the clock time $CT$ is fixed, the true counting intervals will differ for different measurements. This issue can be addressed either by manual correction of the \lt{} or by automatically fixing the \lt{} clock in the multi-channel analyzer (MCA). There are various ``\lt{} clocks'' built into the systems widely used in research and industry. The Gedcke-Hale \lt{} clock is an example of one of the more well-proven, reliable \lt{} clocks.

%------------------------------------------------

Bias can be introduced in a counting measurement by inaccuracy in the \lt{} clock. The degree of this bias appears small in general, and over-correction by the MCA \lt{} clock appears most common in practice. One test of the \lt{} clock performance would be to insert a pulser into the counting system and compare the measured rate with the rate measured directly with a precision frequency counter. As a simple model, the effect of overcorrection (or bias) in the \lt{} clock can be thought of as an error $\delta \tau$ in Eq.\ (\ref{eq:nonextending}) associated with each registered pulse. Then, the perturbation from this \dt{} overcorrection can be estimated as:
\begin{eqnarray}
\delta I &=& \left(\frac{\partial I}{\partial \tau }\right)\delta \tau\\
&=& \left(\frac{I'}{1-\tau I'}\right)^2 \delta \tau = I^2 \delta \tau\;.
\end{eqnarray}
We thus see that the effect of this bias is to overestimate the  counting rate in a manner once again similar to the SID perturbation given by Eq.\ (\ref{sidrate}),
\begin{equation}\label{eq:ltbias}
I + \delta I = I \left(1 + \delta\tau I \right)\;.
\end{equation}
In fact, systematic over or under-correction in the \lt{} clock may lead to an observable behavior identical to the SID behavior.

%----------------------------------------------------------------------------------------
%	SECTION CONTENTS
%----------------------------------------------------------------------------------------
\section{Discussion}
The central observation of the present paper is that the modifications to nuclear decay rates arising from a SID effect are mathematically identical to lowest order with those arising from standard \dt{} effects, as we show explicitly in Eqs. (\ref{sidrate}) and (\ref{eq:pilupexpanded}). Both of these effects produce perturbations from the exponential decay law which have a similar functional dependence on count rates. It then follows that by implementing \dt{} corrections, one could be suppressing or eliminating evidence for the non-exponential behavior expected from the SID effect.

%------------------------------------------------

For the experiment conducted in Ref.\ \cite{Lindstrom1}, a model similar to that discussed in Sec.\ \ref{sec:extending} was adopted to correct for counting losses due to pileup. In particular, the decay data were fitted to Eq.\ (\ref{eq:pileup}) with $I(t) = I(0) \exp\left( -\lambda_0 t\right)$. A $\chi^2$ minimization then yielded  best-fit determinations  of the decay constant $\lambda_0$, the initial activity $I(0)$, and the pileup parameter $\alpha$. We note that if there were a SID effect present in the decay data of Ref.\ \cite{Lindstrom1}, it would manifest itself with a $\xi$-dependence in the pileup parameter $\alpha$, given explicitly by
\begin{equation}\label{eq:alp}
\alpha = \alpha_0 - \frac{2\xi}{I(0)}\;,
\end{equation}
where $\alpha_0$ is meant to denote the true effect from pileup in the absence of SID. What is interesting is that a pulser was also inserted into the system used in Ref.\ \cite{Lindstrom1}, which allows for an alternate method to estimate pileup losses. From Eq.\ (\ref{eq:pulserpileup}), it is evident that the pulser data should allow for a determination of the pileup parameter which lacks the $\xi$-dependence potentially present in Eq.\ (\ref{eq:alp}). It follows that a discrepancy between the two methods for obtaining $\alpha$ may serve as an indication of SID ($\xi \neq 0$).

%------------------------------------------------

\begin{figure}
\includegraphics[width=.9\linewidth]{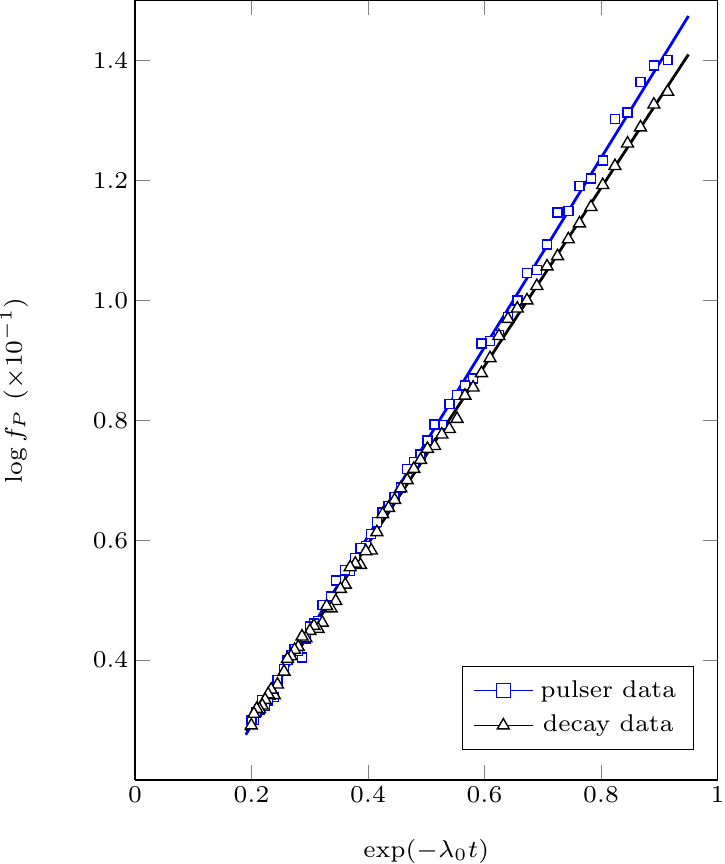}
\caption{Comparison of the correction factors for pileup obtained from the decay data according to the model discussed in the text and those obtained from the pulser data. The prediction from a potential SID effect is to decrease the pileup parameter (the slope) for the black curve.}\label{fig:pill}
\end{figure}

The pulser and decay data for the gold sphere in Ref.\ \cite{Lindstrom1} are plotted in a linearized form of correction factors, similar to Eq.\ (\ref{eq:Fp}), in Fig.\ \ref{fig:pill}. The slopes of the best-fit regressions yield the pileup parameter $\alpha$, and as can be seen in Fig.\ \ref{fig:pill}, the pulser and decay data fit the expected linear models closely. However, Fig.\ \ref{fig:pill} also suggests that the pileup parameter as determined from the pulser data disagrees with $\alpha$ as determined from the decay data, where the discrepancy is approximately $3\sigma$.

%----------------------------------------------------------------------------------------
%	FIGURE 4
%----------------------------------------------------------------------------------------

%------------------------------------------------

Although this observation is suggestive of the presence of SID, it is unclear what other sources may be contributing to this discrepancy. The experiment discussed in Sec.\ \ref{sec:two} raises a similar issue. Although there may be an indication of SID within the data presented, it is also possible that over-correction in the \lt{} clock is responsible for the anomalous behavior. Specifically, a bias in the \lt{} clock according to Eq.\ (\ref{eq:ltbias}) may explain the anomalous behavior described in Sec.\ \ref{sec:two}.

%------------------------------------------------

In principle, the effects of SID can be experimentally distinguished from those of \dt{} in a number of ways. One possible way to resolve this issue is an experiment in which a sample is irradiated to varying degrees of activation and is placed at varying distances from a detector so as to maintain a fixed initial count rate within the detector. In such a configuration the SID effect (which is internal to the sample) should vary, while \dt{} and pileup should remain constant.

%------------------------------------------------

Alternate methods to search for SID exist which eliminate the ambiguity introduced by \dt{} effects. As an example, we note from Fig.\ \ref{fig:cartoon} (and related discussion), that the level of activation achievable in irradiation experiments is affected by the presence of SID. In particular, it is not possible to achieve as high a degree of activation (for $p>0$) for a sample experiencing SID as opposed to one which does not. An experiment could potentially be designed to test for the presence of SID based on this observation. To see this, consider the example of irradiating a sample by neutron activation. The rate of activation $R$ is given by $R = M\sigma J$, where $M$ is the number of target atoms within the sample\footnote{For simplicity, we will neglect the depletion of target atoms (i.e. the time-dependence of $M$) during the activation process.}, $\sigma$ is the cross section for neutron capture, and $J$ is the incident neutron flux. As depicted in Fig.\ \ref{fig:cartoon}, secular equilibrium is reached when the activation rate $R$ equals the decay rate $D$ (which we take to be a positive value). Since $N$ denotes the population of activated atoms, we can write $\dot{N}$ as
\begin{equation}\label{eq:seceqgen}
\dot{N} = R - D\;.
\end{equation}
Secular equilibrium is reached when the population of activated atoms reaches a stationary value, i.e. $R-D = 0$. Since $\dot{N}$ is positive throughout the activation process, we can also conclude that the maximum number of activated atoms $N^*$ is obtained when secular equilibrium is reached. We now compare the two specific cases for secular equilibrium: exponential decay and SID.

%------------------------------------------------

For the standard exponential decay model, Eq.~(\ref{eq:seceqgen}) becomes $R-D = M\sigma J - \lambda_0 N $. Solving for $N(t)$ we find
\begin{equation}\label{eq:29}
N(t) = \frac{M \sigma J}{\lambda_0} \left( 1 - e^{-\lambda_0 t} \right)\;,
\end{equation}
and the condition for secular equilibrium indicates that the maximum number of activated atoms $N^*_{exp}$ is
\begin{equation}\label{eq:Nmaxexp}
N^*_{exp} = \frac{M\sigma J}{\lambda_0}\;,
\end{equation}
which is achieved as $t \rightarrow \infty$. Eqs. (\ref{eq:29}) and (\ref{eq:Nmaxexp}) are valid if $M$ itself is only slowly varying throughout the activation process. The expression for $N(t)$ is useful if we wish to study a sample irradiated for specific period of time which is less than the time required to achieve secular equilibrium.

%------------------------------------------------

In contrast, for the SID model, Eq.~(\ref{eq:seceqgen}) becomes 
\begin{equation}\label{eq:seceqSID}
R-D =  M\sigma J - \frac{\lambda_0 N}{1- pN} \;,
\end{equation}
where we have used Eq.~(\ref{eq:SIDdiff_exact1}) for $D$. The condition for secular equilibrium now implies that the maximum number of activated atoms $N^*_{SID}$ is
\begin{equation}\label{eq:NmaxSID}
N^*_{SID} = \frac{M\sigma J}{\lambda_0 + M\sigma J p}\;.
\end{equation}
This expression can be written in terms of $N^*_{exp}$ to obtain
\begin{equation}\label{eq:NmaxFrac}
\frac{N^*_{SID}}{N^*_{exp}} = \frac{1}{1+p N^*_{exp}}.
\end{equation}
We note from Eq.~(\ref{eq:NmaxSID}) that the maximum number of activated atoms is strictly less than $1/p$, as discussed in Sec.\ \ref{SID_Deadtime_Comparison}. Since $J$ is a measure of the ``power'' of the reactor, taking the limit as $J \rightarrow \infty$ implies that (independent of the reactor) the number of activated atoms can never exceed $1/p$. Of course, since we can never activate more than $M$ atoms (the number of target atoms) either, it follows that
\begin{equation}
N^*_{SID} < \min \left\{ M, 1/p \right\}.
\end{equation}

This is in contrast with the case of standard exponential decay, where $N^*_{exp} <  M$. This discussion leads to a testable prediction: In the presence of SID, a sample cannot be activated to an arbitrary extent. That is, for a large enough sample (where $M > 1/p$) the number of atoms that can be activated is less than some fixed value.

%------------------------------------------------

We conclude this discussion by noting that both the suggestive evidence for SID based on the data presented in Sec.\ \ref{sec:two}, and the analysis of the data of Ref.\ \cite{Lindstrom1} presented here, can be accounted for by a similar vale of $\xi$ in Eq.\ (\ref{eq:SIDdiff}), $\xi \approx 10^{-3}$. Given that these analyses rely on different data and inputs, the fact that they arrive at approximately the same value of $\xi$ supports the inference that a SID effect may be present in these data. Additional support comes from data on annual variations of nuclear decay rates whose amplitudes (when determined as fractional changes to decay rates, similar to $\xi$) are also on the order of $10^{-3}$ (see Fig. 1 of Ref.\ \cite{four}). The indication that anomalies in these deay phenomena may be related to one another may provide additional motivation for the experiments discussed here.

%----------------------------------------------------------------------------------------
%	ACKNOWLEDGMENTS
%----------------------------------------------------------------------------------------
\section*{Acknowledgments}
The authors are deeply indebted to Gregory Downing and Richard Lindstrom for helpful discussions and for their comments on an earlier version of this manuscript. We also wish to thank David J. Sinard whose generous financial support has contributed to this paper.

%----------------------------------------------------------------------------------------
%	REFERENCES
%----------------------------------------------------------------------------------------
\section*{References}
\bibliography{mybibfile}

%----------------------------------------------------------------------------------------
%----------------------------------------------------------------------------------------
%----------------------------------------------------------------------------------------
\end{document}